\documentclass[usegraphicx,usenatbib]{mn2e}
\renewcommand\[{\begin{equation}}
\renewcommand\]{\end{equation}}

\def\xv{\textbf{\emph{x}}}
\def\yv{\textbf{\emph{y}}}
\def\av{\textbf{\emph{a}}}
\def\rv{\textbf{\emph{r}}}
\def\xci{x_i}
\def\yci{y_i}
\def\Phic{\Phi_c}

\def\rhoc{\rho_c}

\def\n{\textbf{\emph{n}}}
\def\i{{\rm i}}

\catcode`\@=11
\def\gsim{\ifmmode{\mathrel{\mathpalette\@versim>}}
    \else{$\mathrel{\mathpalette\@versim>}$}\fi}
\def\lsim{\ifmmode{\mathrel{\mathpalette\@versim<}}
    \else{$\mathrel{\mathpalette\@versim<}$}\fi}
\def\@versim#1#2{\lower 2.9truept \vbox{\baselineskip 0pt \lineskip
    0.5truept \ialign{$\m@th#1\hfil##\hfil$\crcr#2\crcr\sim\crcr}}}
\catcode`\@=12

   \title[Exact density-potential pairs]{Exact density-potential pairs 
          from complex shifted axisymmetric systems}

   \author[Ciotti \& Marinacci]
          {Luca Ciotti and Federico Marinacci \vspace{0.0cm}
           \\ Astronomy Department, University of Bologna, 
                       via Ranzani 1, 40127 Bologna, Italy
          }

\date{Accepted 2008 April 4.  Received 2008 April 1; in original form 2008 February 29}
\usepackage[fleqn]{amsmath}
\usepackage{amsfonts}

\begin{document} 
\maketitle

\begin{abstract} 
In a previous paper the complex-shift method has been applied to self-gravitating spherical systems, producing new analytical axisymmetric density-potential pairs. We now extend the treatement to the Miyamoto-Nagai disc and to the Binney logarithmic halo, and we study the resulting axisymmetric and triaxial analytical density-potential pairs; we also show how to obtain the surface density of shifted systems from the complex-shift of the surface density of the parent model. In particular, the systems obtained from Miyamoto-Nagai discs can be used to describe disc galaxies with a peanut-shaped bulge or with a central triaxial bar, depending on the direction of the shift vector. By using a constructive method that can be applied to generic axisymmetric systems, we finally show that the Miyamoto-Nagai and the Satoh discs, and the Binney logarithmic halo, cannot be obtained from the complex-shift of any spherical parent distribution. As a by-product of this study we also found two new generating functions in closed form for even and odd Legendre polynomials, respectively.

\end{abstract}


\section{Introduction}

For the discussion of many astrophysical problems where gravity is
important, a major difficulty is set by the potential theory, in particular in the applications where the availability of analytical density-potential pairs deviating from spherical symmetry is important. Due both to its relevance and its relation with electrostatic and magnetostatic (see Jackson 1999), it is natural that different techniques have been developed to study this problem (e.g., see Binney \& Tremaine 2008; see also Ciotti \& Bertin 2005; Ciotti \& Giampieri 2007, hereafter CG07, and references therein).

In this paper we focus on a particular but very elegant technique. In fact, in the context of classical electrodynamics  Newman (1973; see also
Newman \& Janis 1965, Newman et al. 1965) considered the case of
the electromagnetic field of a point charge displaced on the imaginary
axis; the remarkable properties of the resulting field have been studied by Carter (1968), Lynden-Bell (2000, 2002,
2004a,b, and references therein; see also Teukolsky 1973;
Chandrasekhar 1976; Page 1976), and Kaiser (2004, and references
therein). This procedure, also known as the \textit{complex-shift method}, was first introduced by Appell (1887; see also Whittaker \& Watson 1950) to the case of the gravitational field of a point mass, and successively used in general relativity (e.g., see Gleiser \& Pullin 1989; Letelier \& Oliveira 1987, 1998; D'Afonseca, Letelier \& Oliveira 2005 and references therein). In the case of classical gravitation, this method has been extended to continuous mass density distributions by CG07, who considered \textit{spherical} parent systems, and obtained new analytical axisymmetric density-potential pairs.

This paper is the natural follow-up of the study of CG07 to \textit{axisymmetric} parent systems, not only to further extend the class of new axisymmetric systems, but also to obtain new triaxial density-potential pairs. In Section 2 we briefly recall the framework developed in CG07, and then we discuss the application of the method to axisymmetric parent models. In Section 3 we describe the new axisymmetric and triaxial systems obtained from the axial and the equatorial shift of the Miyamoto-Nagai disc and the Binney logarithmic halo, while in Section 4 we discuss the projected surface density of the new models. Section 5 summarizes the main results obtained. In Appendix A we discuss the expansion of the complexified potential of a generic assigned density distribution in terms of the shift amplitude and, as a by-product, we found two new generating functions in closed form for even and odd Legendre polynomials, respectively. In Appendix B we finally present the general solution of an inverse problem associated with the complex-shift method.

\section{General considerations}

Here we recall the basic idea of CG07, who extended the complexification of the Coulomb field of a point charge (e.g., see Lynden-Bell 2004b), to the gravitational potential
$\Phi(\xv)$ of a (nowhere negative) density distribution $\rho(\xv)$. From now
on $\xv =(x,y,z)$ indicates the position vector, while
$<\xv,\yv>\equiv\xci\yci$ is the standard inner product over the reals
(repeated index summation convention implied). Let assume that $\rho
(\xv)$ satisfies the Poisson equation
\[
\nabla^2\Phi=4\pi G\rho,
\label{poisson}
\]
and that the associated complexified potential $\Phic$ with shift $\i
\av$ is defined as
\[
\Phic(\xv)\equiv \Phi(\xv -\i\av),
\]
where $\i^2=-1$ is the imaginary unit and $\av$ is a real vector. From the linearity of the coordinate transformation and of the Laplace operator it follows that
\[
\nabla^2\Phic=4\pi G\rhoc,
\]
where 
\[
\rhoc(\xv)\equiv \rho(\xv -\i\av)
\]
is the shifted density; we recall that in this technique 
$||\xv - \i\av ||^{2} = || \xv ||^{2} - || \av ||^{2} - 2\i<\av,\xv>$
 (e.g., see Lynden-Bell 2004b).
By separating the real and imaginary parts of $\Phic$ and
$\rhoc$ obtained from the shift of a known real density-potential pair (or, if simpler, taking the Laplacian of $\Phi_{{\rm c}}$),
one obtains {\it two} new density-potential pairs.
Of course, positivity of the new densities is not assured. In fact
CG07 proved that for a generic parent density $\rho$, the imaginary part $\Im(\rho_{{\rm c}})$ necessarily changes sign because the total mass (and the total gravitational energy) of the complexified density-potential pair coincides with the corresponding quantities of the parent (real) density-potential pair, so that (from the mass invariance) it follows that $\int \Im(\rho_{{\rm c}}) \, d^{3}\xv = 0$. This implies that $\Im(\rho_{{\rm c}})$ cannot be used to describe a gravitating system. On the other hand, either cases can happen for the real part $\Re(\rho_{{\rm c}})$,
depending on the specific seed density and shift vector adopted. For example, while $\Re(\rho_{{\rm c}})$ of the shifted Plummer (1911) and Isochrone (H\'enon 1959) spheres is positive over the whole space when the module of the shift vector is less than a critical value $a_{max}$, $\Re(\rho_{{\rm c}})$ of the shifted singular isothermal sphere has negative regions independently of the adopted $\av \neq 0$ (see CG07).

\subsection{Axisymmetric parent systems}

At variance of the spherical case, the result of the complex shift of an axially symmetric potential $\Phi(R,z)$, where $R = \sqrt{x^{2} + y^{2}}$ is the cilindrical radius, depends on the direction of the shift vector $\mathbf{a}$. For simplicity in the following Sections we consider two special shift directions, namely the axial shift $\mathbf{a} = (0,0,a)$, and the equatorial shift, where without loss of generality we assume $\mathbf{a} = (a_{x},0,0)$; however, some general results for arbitrary shift direction and parent density distributions are postponed to Section 4 and Appendix A. The axially shifted potential and density are
\begin{equation}
\label{phiax}
\Phi_{{\rm c}} = \Phi(R, z_{{\rm c}})\,, \quad\quad\quad
\rho_{{\rm c}} = \rho(R, z_{{\rm c}})\,,
\end{equation}
where 
\begin{equation}
\label{shiftedz}
 z_{{\rm c}}^{2} = z^{2} - a^{2} - 2\i az \,,
\end{equation}
while in the case of equatorial shift
\begin{equation}
\Phi_{{\rm c}} = \Phi(R_{{\rm c}},z)\,,\quad\quad\quad
\rho_{{\rm c}} = \rho(R_{{\rm c}},z)\,,
\end{equation}
where
\begin{equation}
\label{shiftcilrad}
R_{{\rm c}}^{2} = R^{2} - a^{2} - 2\i ax 
\end{equation}
is the shifted cylindrical radius.

\section{Results}

\subsection{The shifted Miyamoto-Nagai disc}

We start by considering the family of models originated by the complex shift of the Miyamoto-Nagai disc (1975, hereafter MN; see also Binney \& Tremanine 2008).
For convenience we use the relative potential $\Psi \equiv -\Phi$  so that the MN density-potential pair can be written as
\begin{equation}
\label{psinorm}
\Psi(R,z) =  \frac{1}{\sqrt{R^{2} + (s + \xi)^{2}}} \equiv \frac{1}{\zeta} \,,
\end{equation}
\begin{equation}
\label{rhonorm}
\rho(R,z) = \left(3 + \frac{7s}{\xi} + \frac{5s^{2}}{\xi^{2}} + \frac{sR^{2} + s^{3}}{\xi^{3}}\right)\frac{\Psi^{5}}{4\pi} \,.
\end{equation}
In the expressions above $\xi \equiv \sqrt{1 + z^{2}},$ all the coordinates are in units of the scale-length $b$, while the relative potential and the density are normalized to $GM/b$ and to $M/ b^{3}$, respectively; $M$ is the total mass of the system. For increasing $s$ the density distribution becomes more and more flat and concentrated, while for $s = 0$ it reduces to the Plummer sphere, for which CG07 has already discussed the complexification and showed that for $a < 0.588$ the real part of the shifted density is nowhere negative. We then expect that the method applied to MN discs will again produce nowhere negative real densities, at least for discs with low flattening.

\subsubsection{The axial shift} 

The axially shifted MN density can be obtained by the direct complexification of eq. (\ref{rhonorm}) as
\begin{equation}
\label{shiftdensity}
\rho_{{\rm c}} = \left(3 + \frac{7s}{\xi_{{\rm c}}} + \frac{5s^{2}}{\xi_{{\rm c}}^{2}} + \frac{sR^{2} + s^{3}}{\xi_{{\rm c}}^{3}}\right) \frac{\Psi_{{\rm c}}^{5}}{4\pi} \,;
\end{equation}
in the following the shift modulus $a$ is intended expressed in units of $b$.
We start from the evaluation of $\xi_{{\rm c}}$. From $\xi_{{\rm c}}^{2} = 1 - a^{2} + z^{2} - 2\i az \equiv ue^{\i\theta}$ we obtain
\begin{equation}
\label{u definition}
u = \sqrt{(1 - a^{2} + z^{2})^{2} + 4a^{2}z^{2}} \,, 
\end{equation}

\begin{equation}
\label{theta definition}
\cos\theta = \frac{1 - a^{2} + z^{2}}{u} \,,\,\,\,\,\,\,\,\,\, \sin\theta = -\frac{2az}{u}\,,
\end{equation}
where in eq. (\ref{u definition}) (and in all analogous cases) the square root is the positive arithmetic operator; thus
\begin{equation}
\label{xic}
\xi_{{\rm c}} = \sqrt{u} e^{\i\theta/2 + ik\pi} \,,\,\,\,\,\,\,\,\, (k = 0, 1) \,.
\end{equation}
Note that $\cos\theta > 0$ for $a < 1$: for simplicity we restrict to this case. In particular, we cut the complex $(u,\theta)$ plane along the negative real axis so that $-\pi < \theta < \pi$. In addition, the natural request that $\xi_{{\rm c}} > 0$ for $z = 0$ 
\begin{figure*}
\vskip -0.3truecm
\hskip -0.2truecm
\includegraphics[height=0.25\textheight,width=1.0\textwidth]{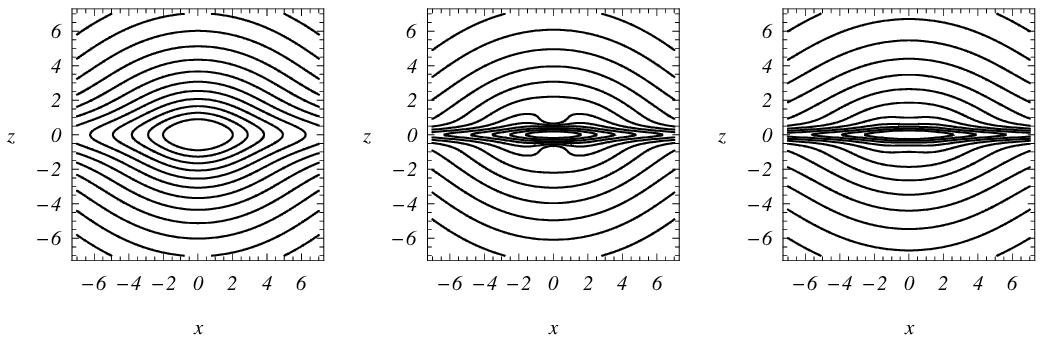}
\end{figure*}
\begin{figure*}
\hskip -0.2truecm
\includegraphics[height=0.25\textheight,width=1.0\textwidth]{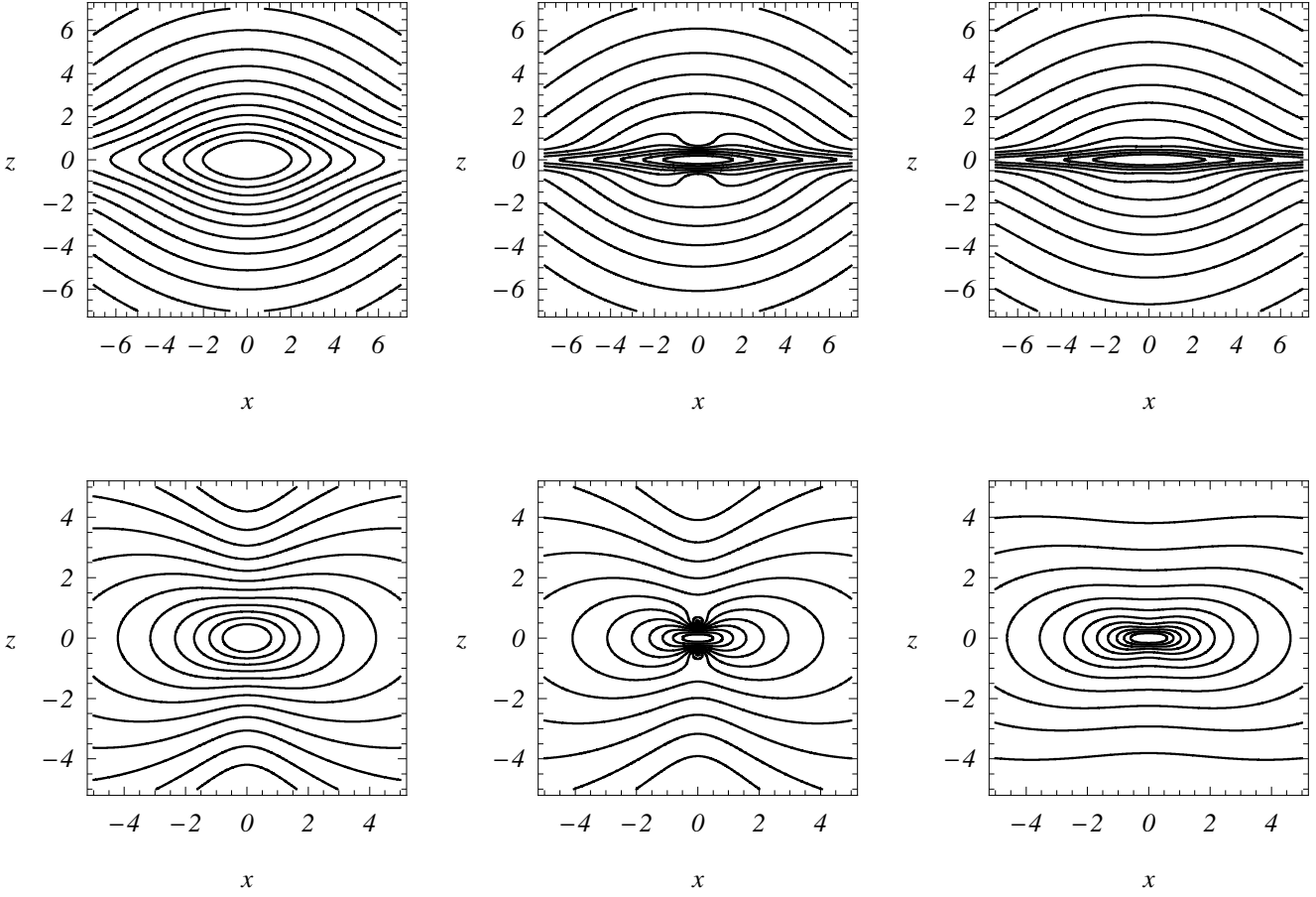}
\caption{Top panels: $\Re(\rho_{{\rm c}})$ of
the axially shifted MN disc with $s = 3/2$. Left and central panels: logarithmically spaced isodensity contours on a meridional plane of the parent ($a = 0$) and of the shifted model with $a = 3/4$, respectively. Top right panel: edge-on surface density for the $a = 3/4$ model.
Bottom panels: $\Re(\rho_{{\rm c}})$ of
the axially shifted Binney logarithmic halo with $q = 0.71$. Left and central panels: logarithmically spaced isodensity contour on a meridional plane of the parent ($a = 0$) and of the shifted model with $a = 1/2$, respectively. Bottom right panel: edge-on surface density for the $a = 1/2$ model. }
\label{isodensity ax MN1}
\end{figure*}
leads to choose $k = 0$. The bisection formulae for $\theta$ are 
\begin{equation}
\label{cossin}
\cos \frac{\theta}{2} = \sqrt{\frac{1 + \cos \theta}{2}} \;,\;\;\;\;\;\;\; \sin \frac{\theta}{2} = \frac{\sin\theta}{2\cos (\theta/2)} \,,
\end{equation} 
where the sign determination is dictated by the interval of variation of $\theta$,
and the real and imaginary parts of $\xi_{{\rm c}}$ are obtained from eqs. (\ref{u definition})-(\ref{cossin}) as
\begin{equation}
\Re(\xi_{{\rm c}}) = \sqrt{\frac{1 - a^{2} + z^{2} + u}{2}} \,, \,\,\,\,\,\,\,\,
\Im(\xi_{{\rm c}}) = - \frac{az}{\Re(\xi_{{\rm c}})}  \,.
\end{equation}

We are now in position to evaluate $\Psi_{{\rm c}} = 1 / \zeta_{{\rm c}}$ from eq. (\ref{psinorm}). By writing $\zeta_{{\rm c}}^{2} = R^{2} + (s + \xi_{{\rm c}})^{2} \equiv v e^{\i\varphi}$, we obtain after some algebra 
\begin{eqnarray*}
 v = \left[\left(R^{2} +s^{2} + u\cos\theta + 2s\sqrt{u}\cos\frac{\theta}{2}\right)^{2}\right. 
\end{eqnarray*}
\begin{equation}
\label{v definition}
 \hspace{0.5cm}  +\,\left. \left(u\sin\theta + 2s\sqrt{u}\sin\frac{\theta}{2}\right)^{2}\right]^{1/2} \,,
\end{equation}

\begin{equation}
\cos\varphi =  \frac{R^{2} +s^{2} + u\cos\theta + 2s\sqrt{u}\cos(\theta/2)}{v} \,,
\end{equation}

\begin{equation}
\label{sin phi definition}
\sin\varphi =  \frac{u\sin\theta + 2s\sqrt{u}\sin(\theta/2)}{v} \,.
\end{equation}
From the positivity of $\cos\theta$ it follows that also $\cos\varphi > 0$, while $\sin\varphi$ changes sign as $\sin\theta$: the formulae for $\cos(\varphi/2)$ and $\sin(\varphi/2)$ are analogous to those in eq. (\ref{cossin}), and  the same considerations followed to determine $\xi_{{\rm c}}$ apply to $\zeta_{{\rm c}}$, so that $\zeta_{{\rm c}} = \sqrt{v} e^{\i\varphi}$ and  
\begin{equation}
\label{mnaxialpsic}
\Psi_{{\rm c}} = \frac{1}{\zeta_{{\rm c}}} = \frac{e^{-\i\varphi/2}}{\sqrt{v}}\,.
\end{equation}
The real and imaginary part of the shifted potential are given by
\begin{equation}
\label{MN axial pot}
\Re(\Psi_{{\rm c}}) = 
\frac{\sqrt{1 - a^{2} + z^{2}  + v + R^{2} + s^{2} + 2s\Re{(\xi_{{\rm c}})}}}{v\sqrt{2}} \,,
\end{equation}
\begin{equation}
\label{eta}
\Im(\Psi_{{\rm c}}) = \frac{az}{v^{2}\Re(\Psi_{{\rm c}})}\left[1 + \frac{s}{\Re(\xi_{{\rm c}})} \right] \,.
\end{equation}
The complexified density can now be easily determined: the 
\begin{figure*}
\vskip -0.3truecm
\hskip -0.2truecm
\includegraphics[height=0.25\textheight,width=1.0\textwidth]{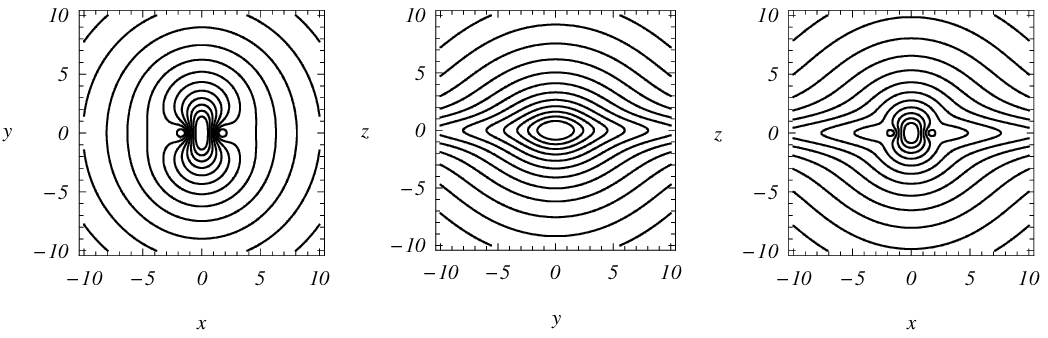}
\end{figure*}
\begin{figure*}
\hskip -0.2cm
\includegraphics[height=0.50\textheight,width=1.0\textwidth]{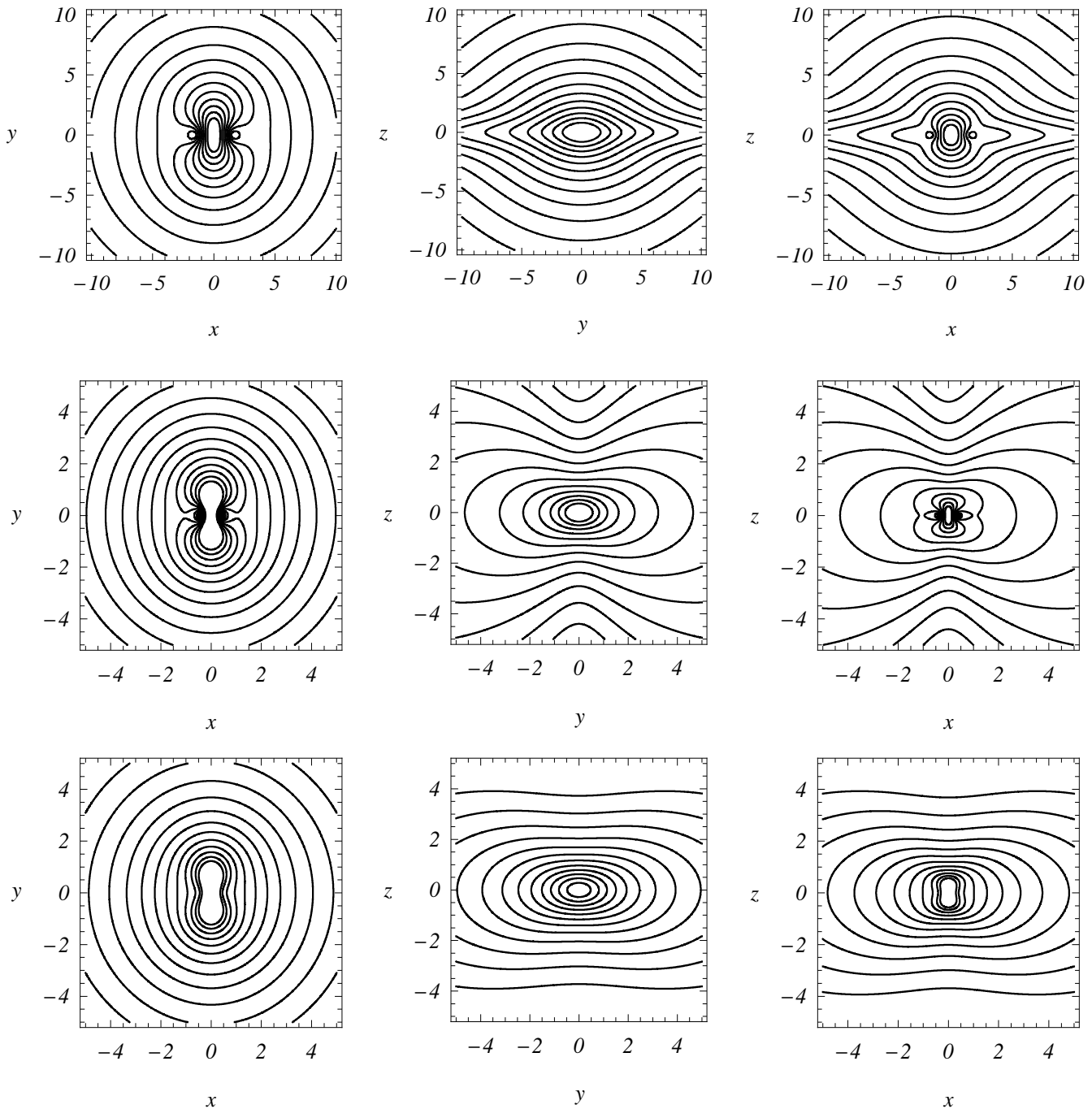}
\caption{Top panels: logarithmically spaced isodensity contours of $\Re(\rho_{{\rm c}})$ of the equatorially shifted MN disc with $s = 3/2$ and $a = 67/40$, in the three orthogonal coordinate planes. The shift is along the $x$-axis and all the lengths are in units of $b$. Central panels: logarithmically spaced isodensity contours of $\Re(\rho_{{\rm c}})$ of the equatorially shifted Binney logarithmic halo with $q = 0.71$ and $a = 0.85$, in the three orthogonal coordinate planes. The shift is along the $x$-axis. Bottom panels: the corresponding projected surface densities of the Binney logarithmic halo.}
\label{isodensity eq MN}
\end{figure*}
simplest way is to substitute $\xi_{{\rm c}}$ (eq. [\ref{xic}] with $k = 0$) and $\Psi_{{\rm c}}$ (eq. [\ref{mnaxialpsic}]) in eq. (\ref{shiftdensity}), and to use eqs. (\ref{v definition})-(\ref{sin phi definition}) and (\ref{MN axial pot})-(\ref{eta}) after separation. The explicit formulae are however quite cumbersome and are not reported here.
In any case, a first check was obtained by comparing term by term the limiting case $s = 0$ with the formulae derived in CG07 for the shifted Plummer sphere. As an additional check, we also compared the Laplacian of $\Re(\Psi_{{\rm c}})$ and $\Im(\Psi_{{\rm c}})$ with the corresponding components of the shifted density, for generic $s$.

As expected, $\Im(\rho_{{\rm c}})$ changes sign crossing the equatorial plane of the system. For what concerns $\Re(\rho_{{\rm c}})$, a numerical exploration reveals that the density becomes negative on the symmetry axis at $z \simeq 0.811$ for $s = 1/2$ and $a \,\gsim\, 0.672$, while the same thing happens at $z \simeq 0.742$ for $s = 3/2$ and $a \,\gsim\, 0.764$. Thus, $a_{max}$ increases for increasing flattening.
In Fig. \ref{isodensity ax MN1} (first and second panels of the top row) we show the isodensity contours of $\Re(\rho_{{\rm c}})$ in a meridional plane for the unshifted model and the model with $a = 3/4$, i.e. a value of the shift parameter near to the critical value. It is apparent how at large distance from the centre the density distribution is almost unaffected by the shift: this is not a pecularity of the present model, but it is a general property of the complex shift, as already recognized by CG07 in the spherical cases, and proved in general in Appendix A. The second feature is the toroidal shape of the density near the centre, with a characteristic depression along the shift axis: such structure is curiously similar to the toroidal shapes of the homeoidally expanded density-potential pairs of Ciotti \& Bertin (2005) and to MOND density-potential pairs (Ciotti, Nipoti \& Londrillo 2006), obtained through potential mapping of homeoidally expanded Newtonian distribution. The resulting density distribution is reminescent of peanut-shaped bulges (e.g., Shaw, Wilkinson \& Carter 1993; Kuijken \& Merryfield 1995; Emsellem \& Arsenault 1997; Bureau \& Freeman 1999; L\"{u}tticke, Dettmar \& Pohlen 2000), and the axially shifted MN disc could then be used as a compact analytical density-potential pair to investigate stellar orbits and gas dynamis in disc galaxies with peanut-shaped bulges (e.g., see Binney \& Petrou 1985; Combes et al. 1990; Patsis, Skokos \& Athanassoula 2002; Quillen 2002; Athanassoula 2005; De Battista et. al 2005).

\subsubsection{The equatorial shift} 

We now consider the equatorial shift of the MN disc and without loss of generality we assume the shift direction  along the $x$ axis. In this case the variable $\xi$ in eq. (\ref{psinorm}) is unaffected by the shift, while the variable $R$ transforms according to eq. (\ref{shiftcilrad}) as $\zeta_{{\rm c}}^{2} = R^{2} + (s + \xi)^{2} - a^{2} - 2\i ax  \equiv v e^{\i\varphi}$,
where
\begin{equation}
v = \sqrt{[R^{2} + (s + \xi)^{2} - a^{2} ]^{2} + 4a^{2}x^{2}} \,,
\end{equation}
\begin{equation}
\cos \varphi = \frac{R^{2} + (s + \xi)^{2} - a^{2}}{v} \,\,\,\,,\,\,\,\,\,\,\, \sin \varphi = - \frac{2ax}{v} \,,
\end{equation}
and
\begin{equation}
\label{zcequat}
\zeta_{{\rm c}} = \sqrt{v} e^{\i\varphi/2 + ik\pi} \,,\,\,\,\,\,\, (k = 0,1)\,.
\end{equation}
For simplicity we restrict to the case $a < 1 + s$, so that $\cos \varphi > 0$ everywhere. Again, we cut the complex $(v, \varphi)$ plane along the negative real axis, and we fix $k = 0$ so that $\zeta_{{\rm c}} > 0$ for $z = 0$. Accordingly, $\cos(\varphi/2)$ and $\sin(\varphi/2)$ are given by identities analogous to eq. (\ref{cossin}), and simple algebra shows that
\begin{equation}
\label{repoteq}
\Re(\Psi_{{\rm c}})  = \frac{\sqrt{v + R^{2} + (s + \xi)^{2} - a^{2}}}{v\sqrt{2}} \,,
\end{equation}
\begin{equation}
\label{impoteq}
\Im(\Psi_{{\rm c}}) = \frac{ax}{v^{2}\Re(\Psi_{{\rm c}})} \,.
\end{equation}
As for the axial shift, also in the case of equatorial shift the the simplest way to obtain the shifted density is by direct complexification and expansion of eq. (\ref{rhonorm}). In particular, the function $\xi$ is unaffected by the shift, and after some work we found
\begin{align}
\Re(\rho_{{\rm c}}) & =  \left[3 + \frac{7s}{\xi} + \frac{5s^{2}}{\xi^{2}} + \frac{s(R^{2} + s^{2} - a^{2})}{\xi^{3}}\right]\frac{\Re(\Psi_{{\rm c}}^{5})}{4\pi} \notag \\
\notag \\ 
\label{rerhoeq}
 \, & +  \frac{asx\,\Im(\Psi_{{\rm c}}^{5})}{2\pi\xi^{3}} \,,
\end{align}
where
\begin{equation}
\Re(\Psi_{{\rm c}}^{5}) = \left[\Re(\Psi_{{\rm c}})^{4} - \frac{10a^{2}x^{2}}{v^{4}} + \frac{5a^{4}x^{4}}{v^{8}\Re(\Psi_{{\rm c}})^{4}}\right]\Re(\Psi_{{\rm c}}) \,,
\end{equation}
and
\begin{equation}
\Im(\Psi_{{\rm c}}^{5}) = \left[5\Re(\Psi_{{\rm c}})^{4} - \frac{10a^{2}x^{2}}{v^{4}} + \frac{a^{4}x^{4}}{v^{8}\Re(\Psi_{{\rm c}})^{4}}\right]\Im(\Psi_{{\rm c}})\,.
\end{equation}
The obtained formulae have been verified by evaluating analytically the Laplacian of eq. (\ref{repoteq}) and comparing the resulting formulae with eq. (\ref{rerhoeq}). In the special limit $s = 0$ the new pair coincides, as expected, with the formulae of CG07 for the shifted Plummer sphere.

For what concerns $\Re(\rho_{{\rm c}})$,
numerical explorations reveal that for $s = 1/2$ the density $\Re(\rho_{{\rm c}})$ becomes negative on the $x$ axis at $x \simeq 1.069$  and $a \,\gsim\, 0.943$, and at $x \simeq 1.563$  and $a \,\gsim\, 1.699$ for $s = 3/2$. The values of $a_{max}$ and $s$ are correlated as in the case of axial shift, in the sense that flatter models are able to support larger shifts.
In the three top panels of Fig. \ref{isodensity eq MN}, we present the isodensity contours of $\Re(\rho_{{\rm c}})$ for $s = 3/2$ and $a$ near its critical value in the three orthogonal coordinate planes.
The effect of the shift along the $x$ direction is apparent. In particular, note that isodense in the top left panel (equatorial plane section) would be \textit{perfectly} circular for the parent unshifted MN disc. The same striction effect is apparent in the top right panel (section on the  meridional plane containing the shift vector). Finally, the top central panel is the meridional section in the plane orthogonal to the shift vector and the density reminds that of the parent MN disc. In fact, it is easy to prove that in this plane the isodense corresponds to that of a standard MN disc where $y^{2}$ is replaced by $y^{2} - a^{2}$.
Thus, the resulting model is a new triaxial density-potential pair, with a triaxial central body smoothly joined to an axisymmetric disc in the external regions. Models of this family, due to their simple analytical nature, could be used to study how central bars influence the orbits of stars and gas in galaxy discs, and to provide the gravitational field needed in hydrodynamical simulations of black hole accretion at the centre of disc galaxies via bar fueling.

\subsection{The shifted Binney logarithmic halo}

We now focus on the Binney (1981) logarithmic halo (see also Binney \& Tremaine 2008)
\begin{equation}
\label{binneypotential}
\Phi = \frac{\ln \left( 1 + m^{2}\right)}{2}  \,, \qquad\qquad m^{2} \equiv R^{2} + \frac{z^{2}}{q^{2}}\,,
\end{equation}
\begin{align}
\rho & = \frac{2\left[1 + R^{2}(1 - q^{2})\right]e^{-4\Phi}+ (2q^{2} - 1)e^{-2\Phi}}{4\pi q^{2}} \notag \\ 
\notag \\
\label{binneydensity}
& = \frac{2\left[q^{2} + z^{2}(1 - q^{-2})\right]e^{-4\Phi} + e^{-2\Phi}}{4\pi q^{2}}  \,,
\end{align}
where all the coordinates are normalized to a scale-lenght $b$, the potential to $v_{0}^{2}$ (the asymptotic equatorial circular velocity), and the density to $v_{0}^{2}/Gb^{2}$
(Evans 1993)\footnote{Note that in eq. (23) of Evans (1993) the quantity $R_{{\rm c}}^{2}$ is missing in the last term and $R_{{\rm c}}^{-4}$ in the dimensional coefficient.}.
The parameter $q \geq 1/\sqrt {2}$ controls the model flattening, and the lower limit is required by the positivity of $\rho$. In the following, we will focus on the real part of the shifted pair, as we already know that the pair $\Im(\rho_{{\rm c}})$-$\Im(\Phi_{{\rm c}})$ cannot be used to describe a gravitating system.

\subsubsection{The axial shift}

From eq. (\ref{shiftedz}) it follows that the axially shifted potential is
\begin{equation}
 \label{shiftedpotbinney}
\Phi_{{\rm c}} =  \frac{1}{2} \ln \left(1 - \frac{a^{2}}{q^{2}} + m^{2} - \frac{2\i az}{q^{2}} \right) \,,
\end{equation}
so that
\begin{equation}
\label{repotbinney}
 \Re(\Phi_{{\rm c}}) = \frac{1}{4} \ln \left[\left(1 - \frac{a^{2}}{q^{2}} + m^{2}\right)^{2} + \frac{4a^{2}z^{2}}{q^{4}}\right] \,.
\end{equation}
The real part of the complexified density can be obtained from the Laplacian of eq. (\ref{repotbinney}), however we use here the first identity in eq. (\ref{binneydensity}). In fact,
as the variable $R$ is unaffected by the axial shift, we just substitute eq. (\ref{shiftedpotbinney}) in the expression for $\rho(R^{2},\Phi)$ obtaining after some work
\begin{align}
\Re(\rho_{{\rm c}}) & = \frac{(2q^{2} - 1)}{4\pi q^{2}} \sqrt{e^{4\Re(\Phi_{{\rm c}})} - \frac{4a^{2}z^{2}}{q^{4}}}\,e^{-4\Re(\Phi_{{\rm c}})}  \notag \\
\notag \\
\label{rerhobinney}
& +\,\frac{1 + R^{2}(1 - q^{2})}{2\pi q^{2}}\left[e^{4\Re(\Phi_{{\rm c}})} - \frac{8a^{2}z^{2}}{q^{4}}\right]e^{-8\Re(\Phi_{{\rm c}})} \,.
\end{align}
As in the other cases, we verified by Laplacian evaluation the correctness of eq. (\ref{rerhobinney}).
Note that solving eq. (\ref{repotbinney}) for the variable $z^{2}$, eq. (\ref{rerhobinney}) can be recast in an algebraic form  $\Re(\rho_{{\rm c}}) = F[R^{2},\Re(\Phi_{{\rm c}})]$, i.e. in a form suitable (in principle) for the recovering of the even part of the two-integrals phase-space distribution function (e.g., see Fricke 1953; Lynden-Bell 1962; Toomre 1982; Hunter \& Quian 1993; Ciotti \& Bertin 2005).
Unfortunately, the presence of (algebraic) irrationalities seems to exclude the possibility of a simple analytical inversion.

As for the axial shift of MN discs, the global positivity of $\Re(\rho_{{\rm c}})$ cannot be guaranteed for a generic value of $a$. Indeed, numerical explorations show that for $q = 1$ (the parent spherical case) the density becomes negative on the simmetry axis of the system at $z \simeq 0.464$ for $a \,\gsim\, 0.808$, and for $q = 0.71$ (the parent maximum flattening) a negative region appears at $z \simeq 0.499$ for $a \,\gsim\, 0.503$. Note that at variance with the MN case, the critical $a$ decreases for parent systems increasingly deviating from spherical symmetry (i.e., with smaller $q$). This is not surprising: in fact, while for MN models the flatter models are more dense on the symmetry axis, the logarithmic halo develops a toroidal shape, which is less able to support the additional density decrease produced by the shift.
In Fig. \ref{isodensity ax MN1} (bottom panels) we show the isodensity contours in a meridional plane of $\Re(\rho_{{\rm c}})$ for the model with a shift parameter near the maximum: the appearence of the critical region on the symmetry axis is evident.

\subsubsection{The equatorial shift}

The equatorial shift of potential (\ref{binneypotential}) is obtained from eq. (\ref{shiftcilrad}), leading to
\begin{equation}
\Phi_{{\rm c}} = \frac{\ln \left(1 - a^{2} + m^{2} - 2\i ax\right)}{2}  \,,
\end{equation}
and
\begin{equation}
\label{binneyeqshiftedpotential}
\Re(\Phi_{{\rm c}}) = \frac{\ln \left[\left(1 - a^{2} + m^{2}\right)^{2} + 4a^{2}x^{2}\right]}{4}  \,.
\end{equation}
Again, the real part of the complexified density can be obtained by direct evaluation of the Laplacian of eq. (\ref{binneyeqshiftedpotential}), but here we use the second identity in eq. (\ref{binneydensity}), so that only the evaluation of $\rho(z^{2},\Phi_{{\rm c}})$ is needed.
The real part of complexified density is then given by
\begin{align}
\Re(\rho_{{\rm c}}) & = \frac{(1 - a^{2} + m^{2})e^{-4\Re(\Phi_{{\rm c}})}}{4\pi q^{2}} \notag \\
\notag \\
 & +\,\frac{[q^{2} + z^{2}(1 - q^{-2})][e^{4\Re(\Phi_{{\rm c}})} - 8a^{2}x^{2}]e^{-8\Re(\Phi_{{\rm c}})}}{2\pi q^{2}} \,.
\end{align}

The positivity of $\Re(\rho_{{\rm c}})$ cannot be guaranteed everywhere for a generic value of $a$ because a negative term is present in its expression. Numerical explorations show that this term dominates over the positive for $q = 1$ at $x \simeq 0.464$ when $a \,\gsim\, 0.808$, being this case coincident with that of the axial shift of the spherical model, and for $q = 0.71$ at $x \simeq 0.372$ when $a \,\gsim\, 0.853$. Note that, in contrast with the axial shift case, $a_{max}$ \textit{decreases} for \textit{increasing} $q$. This is expected from the presence of density lobes on the equatorial plane of the parent distribution.
In Fig. \ref{isodensity eq MN} (central panels) we present the isodensity contours of $\Re(\rho_{{\rm c}})$ in the case $q = 0.71$ and $a = 0.85$, in the three orthogonal coordinate planes. The triaxial nature of the resulting structure is apparent, as the similarity of the density in the central panel with that of the parent system, already mentioned in the MN case.

\section{Surface densities}

So far, we have shown how the complex-shift method can be applied to the case of axisymmetric systems to obtain new analytical density-potential pairs. In this Section we extend the analysis to the projected surface density of the shifted models. In general, the surface density $\Sigma_{{\rm c}}$ of a complexified system projected along the direction of $\n$ (unit vector) is defined as 
\begin{equation}
\label{sigmacdefinition}
\Sigma_{{\rm c}}(\xv_{{\bf \perp}}) \equiv \int^{+\infty}_{-\infty} \rho_{{\rm c}} (\xv_{{\bf \perp}} + l\n) \, dl \,,
\end{equation}
where $l = <\xv,\n>$ is the line-of-sight coordinate, and $\xv_{{\bf \perp}} \equiv \xv - l\n$ is the position vector in the projection plane $<\xv_{{\bf \perp}}, \n> = 0$.
From the linearity of the projection integral, it follows that the real and the imaginary parts of $\Sigma_{{\rm c}}$ are the projections of the real and imaginary parts of the shifted density. Remarkably, $\Sigma_{{\rm c}}$ can be obtained directly by shifting the surface density $\Sigma$ of the parent system with shift $\av_{{\bf \perp}} = \av - <\av, \n>\n$. In fact, $\rho_{{\rm c}} = \rho(\xv - \i\av) = \rho[\xv_{{\bf \perp}} - \i\av_{{\bf \perp}} + (l - \i <\av, \n>)\n ]$, and the integration along $l$ in eq. (\ref{sigmacdefinition}) is unaffected by the change of the origin, as the integration interval extends from $-\infty$ to $+\infty$. Thus, from eq. (\ref{sigmacdefinition}) it follows that
\begin{equation}
\label{Sigmacdef}
 \Sigma_{{\rm c}}(\xv_{{\bf \perp}}) = \Sigma(\xv_{{\bf \perp}} - \i\av_{{\bf \perp}}) \,:
\end{equation}
in particular, the surface density of a model projected along the shift direction ($\av_{{\bf \perp}} = 0$) is real and coincides with the surface density of the parent density distribution.
For simplicity, in this Section we will limit our attention to the face-on and the edge-on projection of the models constructed in Sect. 3.

\subsection{Edge-on projection}

The edge-on projection of an axisymmetric density $\rho(R,z)$ can be written without loss of generality as
\begin{equation}
\label{edgeonproj}
\Sigma(x,z) \equiv \int_{-\infty}^{+\infty} \rho(\sqrt{x^{2} + y^{2}}, z) \, dy \,.
\end{equation}
In the present context we identify two special configurations, i.e. the edge-on projection of an axially shifted system, and of a system shifted equatorially in the direction perpendicular to the line-of-sight so that, according to eq. (\ref{Sigmacdef}),
\begin{equation}
\label{edgeonprojections}
\Sigma_{{\rm c}}(x,z) = \Sigma(x, z - \i a)\,,\,\,\,\,\,\,\,\,\,\,\,\,\,\Sigma_{{\rm c}}(x,z) = \Sigma(x - \i a, z) \,,
\end{equation}
respectively.

As a first example we consider the edge-on projection of the MN disc, for which the surface density (normalized to $M/b^{2}$) can be written as 
\begin{equation}
\label{sigmaYZ}
\Sigma(x,z) =  \left(2 +  \frac{5s}{\xi} + \frac{4s^{2}}{\xi^{2}} + \frac{s^{3} + sx^{2}}{\xi^{3}}\right)\frac{\Lambda^{4}}{2\pi} \,,
\end{equation}
where
\begin{equation}
\Lambda \equiv  \frac{1}{\sqrt{x^{2} + (s + \xi)^{2}}} \,,
\end{equation}
(Ciotti \& Pellegrini 1996). From the general considerations above, it follows that eq. (\ref{sigmaYZ}) is also the edge-on surface density of a MN disc equatorially shifted along the line-of-sight direction.
In the case of axial shift, the complexification is obtained by shifting $\xi\rightarrow\xi_{{\rm c}}$ and $\Lambda\rightarrow\Lambda_{{\rm c}}$ in eq. (\ref{sigmaYZ}), according to the first identity in eq. (\ref{edgeonprojections}). Note that the function $\Lambda$ is the potential evaluated on the projection plane ($y = 0$), so that one can use in eq. (\ref{sigmaYZ}) the quantities $\xi_{{\rm c}}$ and $\Psi_{{\rm c}}$ in their polar form (eqs. [\ref{xic}] and [\ref{mnaxialpsic}], respectively). In the case of shift along $x$, the function $\xi$ is unaffected, while $\Lambda_{{\rm c}}$ is obtained as the equatorially shifted potential (eq. [\ref{repoteq}]) evaluated on the projection plane $y = 0$.

For simplicity in Fig. \ref{isodensity ax MN1} (top right panel) we show only the edge-on case of the axial shift of MN disc. The peanut shaped contours are clearly visible, even though less pronounced than in the spatial section (top central panel), as the natural consequence of projection.

The edge-on projection of the Binney halo is elementary, and the surface density, normalized to $v_{0}^{2}/Gb$, is  
\begin{equation}
\label{binneyedgeon}
 \Sigma(x,z) = \frac{1 + x^{2} + z^{2} + q^{2}}{4q^{2}\, (1 + x^{2} + z^{2}/q^{2})^{3/2}} \;\,\,
\end{equation}
also the complexification is trivial, and the resulting formulae are not reported here.
The edge-on view of the axial shift is given in Fig. \ref{isodensity ax MN1} (bottom right panel), while in Fig. \ref{isodensity eq MN} (bottom central and right panels) we show the edge-on projection for a equatorially shifted model along the line-of-sight (central panel), and perpendicular to it (right panel). Again, the general considerations of Sect. 4 assure that the surface density in the central panel coincides with that of the parent unshifted halo.

\subsection{Face-on projection}

The face-on projected surface density of an axisymmetric density $\rho(R,z)$ is given by
\begin{equation}
\label{sigmaR}
\Sigma(R) \equiv \int_{- \infty}^{+\infty} \rho(R,z) \, dz \,,
\end{equation}
so that in the case of axial shift
\begin{equation}
\label{sigmaRAx}
\Sigma_{{\rm c}}(R)  = \Sigma(R)\,, 
\end{equation}
while in the equatorial shift $\av = \av_{{\bf \perp}}$, and
\begin{equation}
\label{sigmaREq}
\Sigma_{{\rm c}}(R)  = \Sigma(R_{{\rm c}}) \,. 
\end{equation}

As an example we show the face-on surface density of the equatorially shifted Binney logarithmic halo (normalized to $v_{0}^{2}/Gb$)
\begin{equation}
 \Sigma(R) = \frac{q(2 + R^{2})}{4\,(1 + R^{2})^{3/2}} \,,
\end{equation}
where the radius is in units of $b$ and the complexification is obtained from eq. (\ref{shiftcilrad}).
Isodensity contours are represented in Fig \ref{isodensity eq MN} (bottom left panel), for a system with a flattening parameter $q = 0.71$, and shift along the $x$-axis of $a = 0.85$. It is apparent how the spatial deformation is mitigated by the projection.
Unfortunately, the face-on projection of the MN disc cannot be expressed in closed form, so that we do not discuss this case.

\section{Conclusions}

In this paper we have shown that the complex-shift method is able to produce new and explicit density-potential pairs with
finite deviation from spherical symmetry.  In particular, the imaginary part of the complexified density
corresponds to a system of null total mass, while the real component
can be positive everywhere (depending on the original parent density 
distribution and the amount of the complex shift), and so used to describe astrophysical systems.

In a natural follow-up of the preliminary investigation of CG07 (that was restricted to parent spherical models) here we considered the case of axial and equatorial shift of axisymmetric systems. As illustrative examples we considered the cases of the Miyamoto-Nagai disc and Binney logarithmic halo, and we found that the shift leads to new and explicit density-potential pairs, with everywhere positive density (for a shift vector modulus smaller than some limit value). From a qualitative point of view, the shifted densities are more and more similar to the parent distribution at large and larger distances from the centre, and this behaviuor is elucidated in Appendix A by means of the integral representation of the complexified potential. As a by-product of this approach we found two generating functions in closed form for even and odd Legendre polynomials, respectively, that at the best of our knowledge were not previously known.

We also showed how the projected surface density of a generic complexified density can be obtained by shifting the projected density of the parent distribution, thus avoiding the projection of the new density. This would be in general more difficult, due to loss of symmetry as a consequence of the shift action. As an application, we constructed the surface densities of the new models presented. In particular, we found that the edge-on surface densities of the axially shifted Miyamoto-Nagai disc resemble the surface brightness of the so-called peanut-shaped bulges. Also relevant is the case of equatorial shift, leading to a family of simple analytical density-potential pairs axisymmetric at large distances but triaxial in the central regions. Thus, the systems produced by the complex shift of Miyamoto-Nagai discs, could be used as simple analytical models to study gas and orbital dynamics in axisymmetric peanut-shaped galaxies.

In this paper, for simplicity, we restricted the analysis to some simple axisymmetric density-potential pairs, but the class of the systems obtained via the complex-shift method could be extended by complexification of the generalized Miyamoto-Nagai and Satoh density-potential pairs constructed by Vogt \& Letelier (2005, 2007) or by a joint application of the complex-shift and Kelvin inversion to the flat rings model of Letelier (2007).

We conclude by discussing an inverse problem related to the present investigation.
In fact, as shown in CG07, the real part of a complex-shifted  spherical system corresponds to a density-potential pair characterized by axial symmetry. In the course of an interesting discussion, Tim de Zeeuw suggested to investigate wheter the most common axisymmetric density-potential pairs in the literature could be the real part of some complexified spherical system. 
Quite surprisingly, it turns out that it is possible to formulate a simple and constructive approach (presented in Appendix B) able to solve this question in general. In our cases, we were able to prove that the three investigated models, namely the Miyamoto-Nagai and Satoh discs, and the Binney logarithmic halo, cannot be originated by the complex shift of any parent spherical distribution.

\section*{Acknowledgments}
We thank the Referee, Patricio S. Letelier, for useful comments.
L.C. thanks Donald Lynden-Bell for interesting discussions at early stages of this work, and Tim de Zeeuw for having suggested the inversion problem in Appendix B.

\appendix

\section{Integral representation}

We start from the integral representation of the complexified potential
\begin{equation}
\label{intphi}
\Phi_{{\rm c}} \equiv - G \int \frac{\rho(\yv)}{||\rv - \i\av||}\, d^{3} \yv = - \,G \int \rho(\yv)(f_{\Re} + \i f_{\Im})\, d^{3} \yv  \,,
\end{equation}
where $\rv \equiv \xv - \yv$, and $\rho(\yv)$ is the parent density distribution. As in the real case, general informations about the behaviour of the potential can be obtained by expansion in special function (such as Legendre polynomials, e.g., Jackson 1999) of the integral kernel. A first possibility is to use the expansion in Legendre polynomials for $||\rv|| \geq ||\av||$, as given by Lynden-Bell (2004b, eq. [1]) in the case of a point charge:
\begin{equation}
\label{Lynden-Bell}
\frac{1}{||\rv - {\rm i}\av||} \equiv \frac{1}{||\rv||}\sum_{n = 0}^{\infty} {\rm i}^{n} \frac{||\av||^{n}}{||\rv||^{n}} P_{n} (\mu) \,, \,\,\,\,\,\,\,\,\,\,\,\,\,\,\,\, ||\rv||\geq||\av||,
\end{equation}
where $\mu$ is the cosine of the angle between $\av$ and $\rv$.
Here we follow an alternative approach, i.e., first we express in integral form the real and imaginary parts of $\Phi_{{\rm c}}$, and then we expand the resulting kernels.
We define $\rv \equiv \xv - \yv$ and we consider
\begin{equation}
\label{eqA2}
||\rv - \i\av||^{2} = ||\rv||^{2} - ||\av||^{2} - 2i<\av,\rv> \equiv v e^{\i \theta} \,,
\end{equation}
so that
\begin{equation}
\label{eqA3}
 v = \sqrt{(||\rv||^{2} - ||\av||^{2})^{2} + 4<\av,\rv>^{2}} \,,
\end{equation}
and, for $v \neq 0$,
\begin{equation}
\label{eqA4}
 \cos \theta = \frac{||\rv||^{2} - ||\av||^{2}}{v} \,,\,\,\,\,\,\,\,\,\,\,\, \sin \theta = -\frac{2<\av,\rv>}{v} \,. 
\end{equation}
Note that $v$ vanishes only on the ring determined by the intersection of the spherical shell $||\rv|| = ||\av||$ centered on $||\xv||$, with the plane $\mathbb{P}$ containing the point $\xv$ and perpendicular to $\av$ of equation $<\av,\rv> = 0$; on this ring the phase angle $\theta$ is not defined. We call $\mathbb{D}$ the disc $||\rv|| < ||\av||$, with $\rv \in \mathbb{P}$. We now determine the square root $\sqrt{v} e^{i\,\theta /2 + \i k\pi}$. All points $ \rv \in \mathbb{P}$ and outside the disc $\mathbb{D}$ are mapped to points in the $(v,\theta)$ plane with $\sin \theta = 0$ and $\cos \theta = 1$, so that the norm in eq. (\ref{eqA2}) is made positive by fixing $k = 0$ and cutting the complex plane along the negative axis, i.e. $- \pi < \theta < \pi$. Moreover, $\cos (\theta/2) > 0$ for $\rv \notin \mathbb{D}$, and the bisection formulae needed to construct the square root are obtained from eq. (\ref{cossin}), (\ref{eqA3}), (\ref{eqA4})
\begin{equation}
\label{eqA5}
\cos \frac{\theta}{2} = \sqrt{\frac{v + ||\rv||^{2} - ||\av||^{2}}{2v}} \;,
\end{equation}
\begin{equation}
\label{eqA6}
 \sin \frac{\theta}{2} = -\frac{\sqrt{2}<\av,\rv>}{\sqrt{v}\sqrt{v + ||\rv||^{2} - ||\av||^{2}}} \,.
\end{equation}
The only delicate region is the disc $\mathbb{D}$, where
$\cos \theta = -1$ and $\sin\theta = \cos(\theta/2) = 0$: the function $\sin(\theta/2)$ is then discontinuous, changing from $+1$ to $-1$ when crossing the disc in the direction of $\av$.
After some work we arrive at
\begin{align}
\label{eqA7}
\begin{cases}
 f_{\Re} = \dfrac{\sqrt{v + ||\rv||^{2} - ||\av||^{2}}}{\sqrt{2} v} \,, \\ \\
f_{\Im} = {\rm sgn}(\mu) \dfrac{\sqrt{v + ||\av||^{2} - ||\rv||^{2}}}{\sqrt{2} v} \,.
\end{cases}
\end{align}
We now consider the behaviour of eq. (\ref{intphi}) for $||\rv|| > ||\av||$ and $||\rv|| < ||\av||$. It follows that for $t = ||\av||/||\rv|| < 1$
\begin{equation}
\label{eqA8}
 f_{\Re} = \frac{H_{+}(t,\mu)}{||\rv||} \,,\,\,\,\,\,\,\,\,\,\, f_{\Im} = \frac{{\rm sgn}(\mu)H_{-}(t,\mu)}{||\rv||} \,,
\end{equation}
while for $t = ||\rv||/||\av|| < 1$
\begin{equation}
 f_{\Re} = \frac{H_{-}(t,\mu)}{||\av||} \,,\,\,\,\,\,\,\,\,\,\, f_{\Im} = \frac{{\rm sgn}(\mu)H_{+}(t,\mu)}{||\av||} \,,
\end{equation}
where
\begin{align}
\begin{cases}
H_{+}(t,\mu) \equiv 
\dfrac{\sqrt{1 - t^{2} + \sqrt{(1 - t^{2})^{2} + 4t^{2}\mu^{2}}}}{\sqrt{2}\sqrt{(1 - t^{2})^{2} + 4t^{2}\mu^{2}}}, \\ \\
H_{-}(t,\mu) \equiv
\dfrac{\sqrt{t^{2} - 1 + \sqrt{(1 - t^{2})^{2} + 4t^{2}\mu^{2}}}}{\sqrt{2}\sqrt{(1 - t^{2})^{2} + 4t^{2}\mu^{2}}}.
\end{cases}
\end{align}
Two interesting formulae can now be obtained for the functions $H_{+}$ and $H_{-}$.
In fact, separating eq. (\ref{Lynden-Bell}) in its real (even powers) and imaginary (odd powers) parts, and comparing them with eqs. (\ref{eqA8}), it follows that
\begin{align}
\label{Hplus}
\begin{cases}
H_{+}(t,\mu) = \displaystyle{\sum_{n = 0}^{\infty}} (-1)^{n} \, P_{2n}(\mu) \, t^{2n}\,, \\ \\
{\rm sgn}(\mu)H_{-}(t,\mu) = \displaystyle{\sum_{n = 0}^{\infty}} (-1)^{n} \, P_{2n + 1}(\mu) \, t^{2n + 1}\,.
\end{cases}
\end{align}
In other words, $H_{+}$ and $H_{-}$ are generating functions in closed-form for even and odd Legendre polynomials, respectively. We were unable to find this result in the standard literature (e.g., Courant \& Hilbert 1953, Morse \& Feshbach 1953, Erd\'elyi et. al 1955, Abramowitz \& Stegun 1972, Gradshteyn \& Ryzhik 1980, Arfken \& Weber 1995), nor computer algebra systems such as ${\rm Maple}^{\copyright}$ and ${\rm Mathematica}^{\copyright}$ were able to re-sum the series in eq. (\ref{Hplus}) (while able to confirm the expansions).

With the expressions above eq. (\ref{intphi}) becomes
\begin{align}
\label{Phiexpansion}
\Re(\Phi_{{\rm c}}) & = - G \displaystyle{\sum_{n = 0}^{\infty}} (-1)^{n}||\av||^{2n}\int\limits_{||\rv|| \geq ||\av||}  \frac{\rho(\yv)\,P_{2n}(\mu)}{||\rv||^{2n + 1}}  \, d^{3}\yv  \notag \\
& - G \sum_{n = 0}^{\infty} \frac{(-1)^{n}}{||\av||^{2n + 2}} \int\limits_{||\rv|| \leq ||\av||} \!\!\!\!\!\!\rho(\yv)\,P_{2n + 1}(|\mu|)||\rv||^{2n + 1}  \, d^{3}\yv \,.
\end{align}
Now, for a density distribution decreasing sufficiently fast, the integral over $||\xv - \yv|| \leq ||\av||$ is negligible with respect to the other in the far field regions (i.e. $||\xv|| \rightarrow \infty$). In addition, as the even Legendre polynomials are bounded by unity, it results that the leading term in eq. (\ref{Phiexpansion}) is just the potential of the parent system, and this explains why the isodensity contours of the shifted models in the external regions are almost coincident with those of the parent system.

\section{An inversion problem}

The procedure is best illustrated as follows. Let assume that the (unknown) spherically symmetric real function $F(r)$ (potential or density, the argument is identical), is the parent distribution of the investigated axisymmetric system $f(R,z)$, so that $f(R,z) = \Re(F_{{\rm c}})$ for some (unknown) value of the shift parameter $a$. The $z$-shift of amplitude $a$ on the radius $r$ gives
\begin{equation}
\label{SPHCOMPSHIFT}
F_{{\rm c}} = F(r_{{\rm c}}) \,,
\end{equation}
and all the considerations of Appendix A hold, where now, due to the special orientation of $\av$,
$r_{{\rm c}}^{2} = r^{2} - a^{2} - 2iaz = v e^{\i\varphi}$,
\begin{equation}
v = \sqrt{(r^{2} - a^{2})^{2} + 4a^{2}z^{2}}\,,
\end{equation}
\begin{equation}
\cos\theta = \frac{r^{2} - a^{2}}{v} \,,\,\,\,\,\,\,\,\,\,\,\,\,\,\,\,\,\,\,\, \sin\theta = -\frac{2az}{v} \,.
\end{equation}
By evaluating $r_{{\rm c}} = \sqrt{v} e^{\i\theta/2}$, from eqs. (\ref{eqA5})-(\ref{eqA6}) and following comments, it turns out that on the equatorial plane ($z = 0$)
\begin{equation}
\label{phieqplane1}
F_{{\rm c}} =
\begin{cases}
F\left(\sqrt{R^{2} - a^{2}}\right), & \text{$R \geq a$,}\\

 F\left(\pm \i\sqrt{a^{2} - R^{2}}\right), & \text{$R \leq a$,} 
\end{cases}
\end{equation}
where $R$ is the cylindrical radius, and the two signs correspond to the limits $z\rightarrow 0^{-}$ and $z\rightarrow 0^{+}$ for $R \in \mathbb{D}$.
Equation (\ref{phieqplane1}) reveals that on the equatorial plane $F_{{\rm c}}$ is linked to the parent spherical potential by a very simple substitution. In particular, the function $f(R,0)$ for $R \geq a$ gives the whole function $F(r)$ (as the variable $\sqrt{R^{2} - a^{2}}$ spans the entire range $0 \leq r < \infty$ and $F_{{\rm c}}$ coincides with its real part), while $f(R,0)$ for $R \leq a$ is the real part of the map of $F(r)$ for $\rv \in \mathbb{D}$. In practice, for an assigned $f(R,z)$, we construct the putative parent function $F(r) \equiv f(\sqrt{r^{2} + a^{2}},0)$, with $a$ free parameter.
We then shift (with a shift parameter $b$) the spherical candidate, and check if its real part can be made equal to $f(R,z)$ for some choice of $a$ and $b$. In case of impossibility, it follows that the original distribution cannot be obtained from the complex shift of any 
spherical system.

We now apply the considerations above to the MN and the Satoh (1980) discs, and to the Binney logarithmic halo.
Simple algebra shows that for the MN disc, the evaluation of eq. (\ref{psinorm}) for $z = 0$ and the successive substitution $R^{2} \rightarrow r^{2} + a^{2}$ leads to a spherical candidate $F(r)$ given by the potential of a Plummer sphere of scale-lenght $a^{2} + (1 + s)^{2}$. 
A similar result is reached for the Satoh (1980) disc, whose normalized potential is
\begin{equation}
\label{satoh}
\Psi(R,z) =  \frac{1}{\sqrt{R^{2} + z^{2} + s(s + 2\sqrt{1 + z^{2})}}}\,,
\end{equation}
where $s$ is the free parameter controlling the disc flattening (Binney \& Tremaine 2008).
From the substitution $R^{2} = r^{2} + a^{2}$ for $z = 0$ we obtain
again, as spherical candidate $F(r)$, the potential of a Plummer sphere of scale-lenght $a^{2} + s(2 + s)$. Thus, if the MN and Satoh discs are generated by the complex shift of a spherical model, this is a Plummer sphere. However, the complex shift of a Plummer sphere does not generate a MN or a Satoh disc (see CG07), and this close the investigation.
In the case of Binney logarithmic halo (\ref{binneypotential}), the restriction to the equatorial plane and the radial substitution leads to
\begin{equation}
\label{binneysphparent}
F(r) =  \frac{\ln \left(1 + r^{2} + a^{2} \right)}{2}  \,,
\end{equation}
but its complexification with shift $b$ reads
\begin{equation}
\label{shiftedbinneyparent}
\Re(F_{{\rm c}}) =  \frac{\ln\left[\left(1 + r^{2} + a^{2} - b^{2}\right)^{2} + 4b^{2}z^{2}\right]}{4}  \,,
\end{equation}
an expression coincident with  eq. (\ref{binneypotential}) only in the trivial case $q = 1$ and $b = a = 0$. We conclude that also the Binney logarithmic halo cannot be derived from the complex shift of any spherical system.


\begin{thebibliography}{99}

\bibitem{} Abramowitz M., Stegun I.A., 1972,
           Handbook of Mathematical Functions: with Formulas, Graphs, and Mathematical Tables. Dover, New York

\bibitem{} Appell P., 1887,
           Ann. Math. Lpz., 30, 155

\bibitem{} Arfken G.B., Weber H.J., 1995,
           Mathematical Methods for Physicists, 4th Ed. Academic Press, San Diego


\bibitem{} Athanassoula E., 2005,
           MNRAS, 358, 1477

\bibitem{} Binney J., 1981, 
           MNRAS, 196, 455

\bibitem{} Binney J., Petrou M., 1985, 
           MNRAS, 214, 449

\bibitem{} Binney J., Tremaine S., 2008,
           Galactic Dynamics, 2nd Ed. Princeton University Press, Princeton

\bibitem{} Bureau M., Freeman K.C., 1999, 
           AJ, 118, 126


\bibitem{} Carter B., 1968,
           Commun. Math. Phys., 10, 280

\bibitem{} Chandrasekhar S., 1969,
           Ellipsoidal figures of equilibrium. 
           Yale University Press, New Haven

\bibitem{} Chandrasekhar S., 1976,
           Proc. R. Soc. London A, 349, 571


\bibitem{} Ciotti L., Bertin G., 2005, 
           A\&A, 437, 419

\bibitem{} Ciotti L., Giampieri G., 2007,
           MNRAS, 376, 1162 (CG07)

\bibitem{} Ciotti L., Pellegrini S., 1996, 
           MNRAS, 279, 240


\bibitem{} Ciotti L., Nipoti C., Londrillo P., 2006, 
           ApJ, 640, 741

\bibitem{} Combes F., Debbash F., Friendli D., Pfenninger D., 1990, 
           A\&A, 233, 82


\bibitem{} Courant R., Hilbert D.,  1953,
	   Methods of Mathematical Physics. Wiley, New York

\bibitem{} D'Afonseca L.A., Letelier P.S., Oliveira S.R., 2005,
           Class. Quantum Grav., 22, 3803

\bibitem{} De Battista V.P., Carollo C.M., Mayer L., Moore B., 2005,
           ApJ, 628, 678


\bibitem{} Emsellem E., Arsenault R., 1997,
           A\&A, 318, L19

\bibitem{} Erd\'elyi A., Magnus W., Oberhettinger F., Tricomi G., 1955,
           Higher Transcendental Functions. McGraw-Hill, New York

\bibitem{} Evans N.W., 1993,
           MNRAS, 260, 191


\bibitem{} Fricke W., 1952, Astron. Nachr., 280, 193

\bibitem{} Gleiser R., Pullin J., 1989,
           Class. Quantum Grav., 6, 977

\bibitem{} Gradshteyn I.S., Ryzhik I.M., 1980,
           Tables of Integrals, Series and Products, 4th Ed. Academic Press, San Diego


\bibitem{} H\'enon M., 1959,
           Ann. d'Astrophys., 22, 126

\bibitem{} Hunter C., Qian E., 1993,
           MNRAS, 262, 401


\bibitem{} Jackson J.D, 1999,
           Classical Electrodynamics, 3rd Ed. Wiley, New York

\bibitem{} Kaiser G., 2004,
           J. Phys. A: Math. Gen., 37, 8735

\bibitem{} Kellogg O.D., 1953,
           Foundations of potential theory. Dover, New York

\bibitem{} Kuijken K., Merryfield M.R., 1995,
           ApJ, 443, L13

\bibitem{} Letelier P.S., 2007, MNRAS, 381, 1031

\bibitem{} Letelier P.S., Oliveira S.R., 1987,
           J. Math. Phys., 28, 165

\bibitem{} Letelier P.S., Oliveira S.R., 1998,
           Class. Quantum. Grav., 15, 421

\bibitem{} L\"{u}tticke R., Dettmar R.J., Pohlen M., 2000,
           A\&A, 362, 435

\bibitem{} Lynden-Bell D., 1962, 
           MNRAS, 123, 447

\bibitem{} Lynden-Bell D., 2000,
           MNRAS, 312, 301

\bibitem{} Lynden-Bell D., 2002,
           preprint (astro-ph/0207064)

\bibitem{} Lynden-Bell D., 2004a,
           Phys. Rev. D, 70, 104021

\bibitem{} Lynden-Bell D., 2004b,
           Phys. Rev. D, 70, 105017

\bibitem{} Miyamoto M., Nagai R., 1975, 
           PASJ, 27, 533


\bibitem{} Morse P.M., Feshbach H., 1953, 
           Methods of Theoretical Physics. McGraw-Hill, New York


\bibitem{} Newman E.T., 1973,
           J. Math. Phys., 14, 102

\bibitem{} Newman E.T., Janis A.I., 1965,
           J. Math. Phys., 6, 915

\bibitem{} Newman E.T., Couch E.C., Chinnapared K., Exton A., Prakash A.,
           Torrence R., 1965,
           J. Math. Phys., 6, 918

\bibitem{} Page D.N., 1976, 
           Phys. Rev. D, 14, 1509

\bibitem{} Patsis P.A., Skokos Ch., Athanassoula E., 2002, 
           MNRAS, 337, 578

\bibitem{} Plummer H.C., 1911,
           MNRAS, 71, 460

\bibitem{} Quillen A.C., 2002,
           AJ, 124, 722

\bibitem{} Satoh C., 1980,
           PASJ, 32, 41

\bibitem{} Shaw M., Wilkinson A., Carter D., 1993,
           A\&A, 268, 511

\bibitem{} Teukolsky S., 1973,
           ApJ, 185, 635

\bibitem{} Toomre A., 1982,
           ApJ, 259, 535

\bibitem{} Vogt D., Letelier P.S., 2005,
           PASJ, 57, 871

\bibitem{} Vogt D., Letelier P.S., 2007,
           PASJ, 59, 319

\bibitem{} Whittaker E.T., Watson G.N., 1950,
           A course of modern analysis. Cambridge University Press, Cambridge



\end{thebibliography}
\end{document}